\documentclass[12pt]{article}
  \usepackage{amsfonts}
  \usepackage{amsmath}
\usepackage{amssymb}
\usepackage{amscd}
\usepackage{graphicx}
\begin{document}

~~
\bigskip
\bigskip
\begin{center}
{\Large {\bf{{{Photoelectric effect for twist-deformed space-time}}}}}
\end{center}
\bigskip
\bigskip
\bigskip
\begin{center}
{{\large ${\rm {Marcin\;Daszkiewicz}}$}}
\end{center}
\bigskip
\begin{center}
\bigskip

{ ${\rm{Institute\; of\; Theoretical\; Physics}}$}

{ ${\rm{ University\; of\; Wroclaw\; pl.\; Maxa\; Borna\; 9,\;
50-206\; Wroclaw,\; Poland}}$}

{ ${\rm{ e-mail:\; marcin@ift.uni.wroc.pl}}$}

\end{center}
\bigskip
\bigskip
\bigskip
\bigskip
\bigskip
\bigskip
\bigskip
\bigskip
\bigskip
\begin{abstract}
In this article, we investigate the impact of twisted space-time on the photoelectric effect, i.e., we derive the $\theta$-deformed threshold frequency. In such a way we indicate that the space-time noncommutativity strongly enhances the photoelectric process.
\end{abstract}
\bigskip
\bigskip
\bigskip
\bigskip
\eject

The suggestion to use noncommutative coordinates goes back to
Heisenberg and was firstly  formalized by Snyder in \cite{snyder}.
Recently, there were also found formal  arguments based mainly  on
Quantum Gravity \cite{2}, \cite{2a} and String Theory models
\cite{recent}, \cite{string1}, indicating that space-time at the Planck
scale  should be noncommutative, i.e., it should  have a quantum
nature. Consequently, there appeared a lot of papers dealing with
noncommutative classical and quantum  mechanics (see e.g.
\cite{mech}, \cite{qm}) as well as with field theoretical models
(see e.g. \cite{prefield}, \cite{field}), in which  the quantum
space-time is employed.

In accordance with the Hopf-algebraic classification of all
deformations of relativistic \cite{class1} and nonrelativistic
\cite{class2} symmetries, one can distinguish three basic types
of space-time noncommutativity (see also \cite{nnh} for details):\\
\\
{ \bf 1)} Canonical ($\theta^{\mu\nu}$-deformed) type of quantum space \cite{oeckl}-\cite{dasz1}
\begin{equation}
[\;{ x}_{\mu},{ x}_{\nu}\;] = i\theta_{\mu\nu}\;, \label{noncomm}
\end{equation}
\\
{ \bf 2)} Lie-algebraic modification of classical space-time \cite{dasz1}-\cite{lie1}
\begin{equation}
[\;{ x}_{\mu},{ x}_{\nu}\;] = i\theta_{\mu\nu}^{\rho}{ x}_{\rho}\;,
\label{noncomm1}
\end{equation}
and\\
\\
{ \bf 3)} Quadratic deformation of Minkowski and Galilei  spaces \cite{dasz1}, \cite{lie1}-\cite{paolo}
\begin{equation}
[\;{ x}_{\mu},{ x}_{\nu}\;] = i\theta_{\mu\nu}^{\rho\tau}{
x}_{\rho}{ x}_{\tau}\;, \label{noncomm2}
\end{equation}
with coefficients $\theta_{\mu\nu}$, $\theta_{\mu\nu}^{\rho}$ and  $\theta_{\mu\nu}^{\rho\tau}$ being constants.\\
\\
Moreover, it has been demonstrated in \cite{nnh}, that in the case of the
so-called N-enlarged Newton-Hooke Hopf algebras
$\,{\mathcal U}^{(N)}_0({ NH}_{\pm})$ the twist deformation
provides the new  space-time noncommutativity of the
form\footnote{$x_0 = ct$.},\footnote{ The discussed space-times have been  defined as the quantum
representation spaces, so-called Hopf modules (see e.g. \cite{oeckl}, \cite{chi}), for the quantum N-enlarged
Newton-Hooke Hopf algebras.}
\begin{equation}
{ \bf 4)}\;\;\;\;\;\;\;\;\;[\;t,{ x}_{i}\;] = 0\;\;\;,\;\;\; [\;{ x}_{i},{ x}_{j}\;] = 
if_{\pm}\left(\frac{t}{\tau}\right)\theta_{ij}(x)
\;, \label{nhspace}
\end{equation}
with time-dependent  functions
$$f_+\left(\frac{t}{\tau}\right) =
f\left(\sinh\left(\frac{t}{\tau}\right),\cosh\left(\frac{t}{\tau}\right)\right)\;\;\;,\;\;\;
f_-\left(\frac{t}{\tau}\right) =
f\left(\sin\left(\frac{t}{\tau}\right),\cos\left(\frac{t}{\tau}\right)\right)\;,$$
$\theta_{ij}(x) \sim \theta_{ij} = {\rm const}$ or
$\theta_{ij}(x) \sim \theta_{ij}^{k}x_k$ and  $\tau$ denoting the time scale parameter
 -  the cosmological constant. Besides, it should be  noted, that the  above mentioned quantum spaces {\bf 1)}, { \bf 2)} and { \bf 3)}
 can be obtained  by the proper contraction limit  of the commutation relations { \bf 4)}\footnote{Such a result indicates that the twisted N-enlarged Newton-Hooke Hopf algebra plays a role of the most general type of quantum group deformation at nonrelativistic level.}.

In this article we investigate the impact of quantum space-times (\ref{nhspace}) on the photoelectric process described by the following equation
\cite{photo}
\begin{equation}
K = \hbar\omega - W\;, \label{equation}
\end{equation}
where $K$, $\hbar\omega$ and $W$ denote the kinematic energy of electron, energy quanta of light and work function, respectively. To this end, we assume
that photons emitted by transplanckian (noncommutative) source are described by the nonrelativistic oscillator model \cite{daszoscy} defined
on the following twist-deformed N-enlarged Newton-Hooke phase space
\begin{eqnarray}
&&[\;\hat{ x}_{1},\hat{ x}_{2}\;] = 2if_{\kappa}({t})\;\;\;,\;\;\;
[\;\hat{ p}_{1},\hat{ p}_{2}\;] = 2ig_{\kappa}({t})\;,\label{rel1}\\
&&[\;\hat{ x}_{i},\hat{ p}_{j}\;] =
i\hbar \delta_{ij}\left[1+ f_{\kappa}({t})g_{\kappa}({t})/\hbar^2\right]\;, \label{rel2}
\end{eqnarray}
with an arbitrary function $g_{\kappa}({t})$\footnote{Essentially, we should  consider Maxwell Field Theory
defined on quantum space (\ref{nhspace}). However, its construction seems to be quite difficult and for this reason
here we consider only toy-model in which  oscillations of emitted light are described by the nonrelativistic and
first quantized noncommutative oscillator model \cite{daszoscy}.}. 
Then the corresponding Hamiltonian operator is given by
\begin{eqnarray}
\hat{H} = \frac{1}{2m}\left({\hat{{p}}_1^2}+{\hat{{p}}_2^2} \right) +
\frac{1}{2}m\omega^2 \left({\hat{{x}}_1^2}+{\hat{{x}}_2^2} \right) \;. \label{2dhn}
\end{eqnarray}
with $m$ and $\omega$ denoting the mass and frequency of a particle, respectively.
In terms of commutative  variables $({ x}_i, { p}_i)$, which correspond to the low-energy observer, it takes the form\footnote{The operators $({ x}_i, { p}_i)$ satisfy
$[\;{ x}_{i},{ x}_{j}\;] = [\;{ p}_{i},{ p}_{j}\;] = 0$,
$[\;{ x}_{i},{ p}_{j}\;] = i\hbar\delta_{ij}$ and
describe, for example,  the surface of metal in a typical laboratory room.}
\begin{eqnarray}
\hat{H} = \hat{H}(t) =
\frac{1}{2M(t)}\left({{{p}}_1^2}+{{{p}}_2^2} \right)  +
\frac{1}{2}M(t)\Omega^2(t)\left({{{x}}_1^2}+{{{x}}_2^2} \right)
- S(t)L\;, \label{2dh1}
\end{eqnarray}
where
\begin{eqnarray}
&&L = x_1p_2 - x_2p_1\;, \\
&&1/M(t) = 1/m +m\omega^2 f_{\kappa}^2(t)/\hbar^2 \;,\\
&&\Omega(t) = \Omega_f(\omega) = \sqrt{\left(1/m
+m\omega^2 f_{\kappa}^2(t)/\hbar^2 \right)\left(m\omega^2
+ g_{\kappa}^2(t)/(\hbar^2m)\right)}\;,
\end{eqnarray}
and
\begin{equation}
S(t)=m\omega^2f_{\kappa}(t)/\hbar +g_{\kappa}(t)/(\hbar m)\;.
\end{equation}
The corresponding energy spectrum can be find with the use of time-dependent creation/anni-hilation  operator procedure and
it looks as follows
\begin{equation}
E_{n_+,n_-}(t) = \hbar\Omega_{+}(t) \left[n_+ + \frac{1}{2}\right] +
\hbar\Omega_{-}(t) \left[n_- + \frac{1}{2}\right]\;\;;\;\;n_\pm = 0, 1, 2, \ldots\;, \label{eigenvalues}
\end{equation}
with frequencies 
\begin{eqnarray}
\Omega_{\pm}(t)&=&\Omega(t)\mp S(t)\;.\label{nn}
\end{eqnarray}
Besides, one can observe that for functions $f_{\kappa}({t})$ and $g_{\kappa}({t})$
such that
\begin{eqnarray}
f_{\kappa}(t) = -{g_{\kappa}(t)}/({\omega^2m^2}) \;,\label{condition}
\end{eqnarray}
we have
\begin{equation}
E_{n_+,n_-}(t) = \hbar\Omega(t) \left[n_+ + \frac{1}{2}\right] +
\hbar\Omega(t) \left[n_- + \frac{1}{2}\right] \;. \label{landau}
\end{equation}
It means that the spectrum (\ref{eigenvalues}) becomes isotropic and the corresponding energy quanta takes the form\footnote{Due to the isotropy of spectrum
 (\ref{landau}) we consider further excitations only in one direction.}
\begin{equation}
E_{quanta}=E_{n+1}-E_n = \hbar \Omega_{f}(\omega)\;. \label{quanta}
\end{equation}
Particularly,
for the canonical deformation $f_{\kappa}(t) = \kappa = \theta = {\rm const}$ we get
\begin{equation}
E_{n_+,n_-,\theta} = \hbar\Omega_{\theta} \left[n_+ + \frac{1}{2}\right] +
\hbar\Omega_{\theta} \left[n_- + \frac{1}{2}\right] \;, \label{landauek}
\end{equation}
with a constant frequency
\begin{equation}
\Omega_{\theta} = \Omega_{\theta}(\omega) = m\omega\left({1}/{m}
+{m\omega^2}\theta^2/{\hbar^2}\right)
\;, \label{stalacze}
\end{equation}
for which $\lim_{\theta \to 0}\Omega_{\theta} = \omega$.

As mentioned above, in the hamiltonian function (\ref{2dhn}), the frequency $\omega$
corresponds to the frequency of emitted light described in terms of noncommutative variables $(\hat{ x}_i, \hat{ p}_i)$ associated with the
Planck scale. However,
the formula (\ref{quanta}) gives the corresponding energy quanta (obviously different than $\hbar \omega$) in terms of
commutative variables $({ x}_i, { p}_i)$,  i.e., it describes the deformed energy of a single photon which is detected by low-energy observer. It is simply the
energy quanta which should  be detected on a surface of metal in a typical laboratory room. In other words the formula (\ref{quanta}) gives
the energy of photons emitted for example by transplanckian (noncommutative) astrophysical sources which arrive to (commutative) Earth.

Consequently,  in our further
analysis, we exchange in (\ref{equation}) the quanta $\hbar\omega$ by the deformed ones $\hbar \Omega_{f}(\omega)$  such that
\begin{equation}
K = \hbar \Omega_{f}(\omega) - W =  \hbar m\omega\left({1}/{m}
+{m\omega^2}f^2_{\kappa}(t)/{\hbar^2}\right) - W\;. \label{defequation}
\end{equation}
One can see that the main difference between (\ref{equation}) and (\ref{defequation}) concerns the shape of the function $K(\omega)$.
In the first (undeformed) case it remains linear in the frequency $\omega$ while for the second process it forms the third degree polynomial.
Next one can ask for so-called threshold quanta, i.e., for such an energy portion for which the frequency $\omega_{tr}$ satisfies
\begin{equation}
\hbar\omega_{tr} - W = 0\;, \label{trequation}
\end{equation}
and
\begin{equation}
\hbar \Omega_{f}(\omega_{tr}) - W = 0\;, \label{trdefequation}
\end{equation}
respectively. The solution of (\ref{trequation}) with respect to the frequency $\omega_{tr}$ seems to be trivial
\begin{equation}
\omega_{tr} = \frac{W}{\hbar}\;. \label{solution}
\end{equation}
In the case of equation (\ref{trdefequation}) the situation is more complicated. However, one can find its three roots - two of them are complex
while the third one remains real; it looks as follows
\begin{eqnarray}
\omega_{tr} (f_{\kappa}(t)) &=&
\frac{\left(\sqrt{3} \sqrt{4 \hbar^6 m^6 f_{\kappa}^6(t)+27 \hbar^2 m^8 f_{\kappa}^8(t) W^2}+9 \hbar m^4 f_{\kappa}^4(t) W\right)^{1/3}}{(2^{1/3} 3^{2/3} m^2 f_{\kappa}^2(t))} +\nonumber\\
&-&\frac{\left(\left(\frac{2}{3}\right)^{1/3} \hbar^2\right)}{\left(\sqrt{3} \sqrt{4 \hbar^6 m^6 f^6_{\kappa}(t)+27
  \hbar^2 m^8 f_{\kappa}^8(t) W^2}+9 \hbar m^4 f_{\kappa}^4(t) W\right)^{1/3}}\;. \label{defsolution}
\end{eqnarray}
The solutions (\ref{solution}) and (\ref{defsolution}) define the threshold frequencies for processes (\ref{equation}) and (\ref{defequation}), respectively.

Let us now turn to the simplest (canonical) deformation of the phase space (\ref{rel1}), (\ref{rel2}) such (as already mentioned) that
\begin{equation}
f_{\kappa}(t) = \theta = {\rm const}\;. \label{cc1}
\end{equation}
Then, in accordance with the relation (\ref{condition}) we have
\begin{equation}
g_{\kappa}(t) = -\theta\omega^2 m^2 \;. \label{cc2}
\end{equation}
Consequently, in such a case, due to the formula (\ref{stalacze}), equation (\ref{defequation}) takes the form
\begin{equation}
K = \hbar \Omega_{\theta}(\omega) - W =  \hbar m\omega\left({1}/{m}
+{m\omega^2}\theta^2/{\hbar^2}\right) - W\;, \label{thetaequation}
\end{equation}
while the corresponding threshold frequency $\omega_{tr}$ is equal to
\begin{eqnarray}
\omega_{tr} (\theta) &=&
\frac{\left(\sqrt{3} \sqrt{4 \hbar^6 m^6 \theta^6+27 \hbar^4 m^8 \theta^8 W^2}+9 \hbar^2 m^4 \theta^4 W\right)^{1/3}}{(2^{1/3} 3^{2/3} m^2 \theta^2)} +\nonumber\\
&-&\frac{\left(\left(\frac{2}{3}\right)^{1/3} \hbar^2\right)}{\left(\sqrt{3} \sqrt{4 \hbar^6 m^6 \theta^6+27
  \hbar^4 m^8 \theta^8 W^2}+9 \hbar^2 m^4 \theta^4 W\right)^{1/3}}\;. \label{thetasolution}
\end{eqnarray}
Obviously, for the deformation parameter $\theta$ approaching zero we should reproduce from (\ref{thetasolution}) the standard relation (\ref{solution}). Besides
we have (see Figure 1 and 2)
\begin{equation}
\lim_{\theta \to \infty} \omega_{tr} (\theta) = 0\;, \label{thetalimit}
\end{equation}
which means that in our treatment the canonical noncommutativity strongly enhances the photoelectric effect.

\section*{Acknowledgments}
The author would like to thank J. Lukierski for valuable discussions.
 This paper has been financially  supported  by Polish
NCN grant No 2014/13/B/ST2/04043.

\eject

$~~~~~~~~~~~~~~~~~~$
\\
\\
\\
\\
\\
\begin{figure}[htp]
\includegraphics[width=\textwidth]{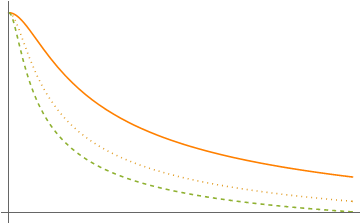}
\caption{The shape of the threshold frequency  $\omega_{tr}(\theta)$ for the three different values of
 the parameter $m$: $m=1$ (continuous line), $m=2$ (dotted line) and $m=3$ (dashed line). In all
three cases we fix the work function $W=1$.}\label{grysunek1}
\end{figure}
\begin{figure}[htp]
\includegraphics[width=\textwidth]{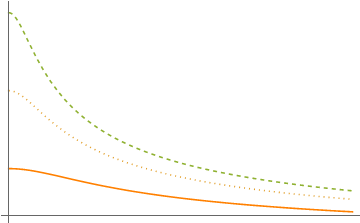}
\caption{The shape of the threshold frequency  $\omega_{tr}(\theta)$ for the three different values of
the work function: $W=1$ (continuous line), $W=2$ (dotted line) and $W=3$ (dashed line). In all
three cases we fix the parameter $m=1$.}\label{grysunek2}
\end{figure}

\end{document}